\documentclass[notoc]{JHEP}
\newcommand{\no}{\nonumber\\}
\newcommand{\be}{\begin{equation}}
\newcommand{\ee}{\end{equation}}

\newcommand{\ba}{\begin{eqnarray}}
\newcommand{\ea}{\end{eqnarray}}
\newcommand{\ci}[1]{\cite{#1}}
\newcommand{\bi}[1]{\bibitem{#1}}
\newcommand{\la}[1]{\label{#1}}

\def\gl#1{(\ref{#1})}
\def\tr#1{\mbox{\rm tr}\left(#1\right)}

\title{Vector mesons in the Extended\\ Chiral Quark Model}

\author{A. A. Andrianov\thanks{Permanent address:
Department of Theoretical
Physics, Sankt-Petersburg State University,
198904 Sankt-Petersburg,  Russia. E-mail: andrian@ecm.ub.es
} and D. Espriu\thanks{E-mail: espriu@ecm.ub.es},\\  Departament 
d'Estructura i Constituents de la Mat\`eria,\\
 Universitat de Barcelona,\\
Diagonal, 647, 08028 Barcelona, Spain,}

\abstract{ We extend our previous formulation of low-energy QCD in
terms of an effective lagrangian containing operators of
dimensionality
$d\le 6$ constructed with pseudoscalars and quark fields, describing physics
below the scale of chiral symmetry breaking. We include in this paper the
vector and axial-vector channels.
We follow closely the Extended
Chiral Quark Model approach and consistently
work in the large-$N_c$ and leading log approximation and take into account
the constraints from chiral symmetry and chiral symmetry restoration.
The optimal fit of all parameters
gives further support to a heavy scalar meson with a
mass $\sim 1$ GeV and a value of the axial pion-quark coupling
constant $g_A \simeq 0.55$ to 0.66, depending on some assumptions
concerning the Weinberg sum rules.}

\preprint{UB-ECM-PF 99/11\\
June 1999\\
hep-ph/9906459}
\begin{document}

\section{ Introduction of the Extended Chiral Quark Model}
\hspace*{3ex}The purpose of this paper is to analyze 
the physics of low-lying scalar, pseudoscalar,
vector and axial-vector mesons
in the framework of  the  Extended Chiral Quark Model (ECQM) 
which was introduced in \ci{aet}.
We shall see that this model of low-energy QCD is sufficiently general and
robust to allow for a good description of the vector and axial-vector
channels
when the model introduced in \ci{aet} is suitably modified to make room for
spin 1 resonances. On the other hand, the addition of vector and
axial-vector resonances constraints
the parameters of ECQM substantially if
experimental data are to be fit consistently.
Special emphasis is put on
the axial pion-quark coupling constant\ci{1} $g_A$ 
whose value happens to be crucial
to obtain phenomenologically acceptable values of meson masses and
coupling constants. This effective coupling was taken as
a free input in the first version of the ECQM.

In a sense
the ECQM lagrangian ${\cal L}_{ECQM}$ systematically extends both the Chiral
Quark (CQM) \ci{1,2} and
Nambu-Jona-Lasinio (NJL) models \ci{3} (see reviews \ci{310}-\ci{bij} and
references therein) and is based on the non-linear realization of chiral
symmetry\footnote{There exist also extensions based on linear realization
of chiral symmetry\ci{7,pp,71}.}. Its inspiration is drawn from Wilson's
renormalization-group ideas as well as the general principles of locality
and gauge and chiral symmetry. The basic idea is to use the degrees
of freedom which are relevant at each energy scale. It is therefore built in
terms of colored current
quark fields $\bar q_i(x), q_i(x)$ with momenta restricted to be below the
chiral symmetry breaking (CSB) scale $\Lambda_{CSB} \sim
1.3$ GeV,
and colorless chiral fields
$U(x) = \exp \left(i\pi(x)/ F_0\right)$  
which are  $SU(N_F)$ matrices
with generators $ \pi \equiv \pi^a T^a$, $a=1,..., N_F^2-1$, and which,
naturally, only appear below $\Lambda_{CSB}$. The quarks can be then
endowed with a relatively large `constituent' mass without
manifestly breaking chiral symmetry. The information
on modes with frequencies larger than $\Lambda_{CSB}$ is contained
in the coefficients of the effective lagrangian, and used as boundary
conditions (at scale $\Lambda_{CSB}$) of the renormalization-group
equations. In addition,
some residual gluon interactions remain below $\Lambda_{CSB}$. These
residual gluon interactions (much diminished after the
explicit inclusion of pions and other resonances) make the model
still confining and thus provide the two point functions with the
correct analytic structure. After a further integration of the heavy
`constituent'
quarks one is left with an effective lagrangian written in terms of
purely physical degrees of freedom. In this lagrangian the net effect
of the residual gluon interactions is to contribute to
the actual value of the coefficients. In fact, they do so by an
amount that, while important, should not be the dominant one.

As in the previous paper we shall restrict ourselves to the case
$N_F = 2$. We add also external
vector,$\bar V_\mu (x)$,
axial-vector, $\bar A_\mu (x)$, scalar, $\bar S (x)$ and pseudoscalar, 
$\bar P (x)$ sources
in order to compute the correlators of corresponding quark currents
and to analyze meson properties. These external
fields are induced by a minimal coupling in the QCD Dirac operator (in
Euclidean notations)
\be
 \hat D \equiv i  \gamma_\mu(\partial_\mu+ \bar V_\mu  + \gamma_5\bar A_\mu) +
 i (\bar S + i \gamma_5 \bar P), \la{exts}
\ee
where $\langle S\rangle = m_q$,  the diagonal matrix of current
quark masses.

The low-energy effective lagrangian
 ${\cal L}_{ECQM}$ must be invariant
under simultaneous left and right $SU(2)$  rotations,
$\Omega_R (x),\, \Omega_L (x)$,
of quark, chiral and external fields \ci{1,2}
\ba
 U &\rightarrow& \Omega_R U \Omega^+_L,\quad
q_L \equiv \frac12 (1 + \gamma_5) q \rightarrow \Omega_L q_L,\quad
q_R  \equiv \frac12 (1 - \gamma_5) q \rightarrow   \Omega_R q_R,\no
\bar L_{\mu} &\equiv&\bar V_\mu +\bar A_\mu  \rightarrow
\Omega_L \bar L_{\mu}  \Omega^+_L + 
\Omega_L \partial_\mu \Omega^+_L ,\no
\bar R_{\mu} &\equiv& \bar V_\mu -
\bar A_\mu \rightarrow \Omega_R \bar R_{\mu} \Omega^+_R + 
\Omega_R \partial_\mu \Omega^+_R ,\no
\bar{\cal M} &\equiv&   \bar S + i\bar P \rightarrow \Omega_R \bar{\cal M}
\Omega^+_L.
\la{trans}
\ea
It is convenient to introduce the `rotated',`dressed' or `constituent' quark
fields \ci{1}
\be
Q_L \equiv \xi q_L,\qquad Q_R \equiv \xi^\dagger q_R,\qquad  \xi^2\equiv U,
\la{chv}
\ee
which transform nonlinearly under  $ SU_L(2) \bigotimes SU_R(2) $
but identically for left and right quark components
\be
\xi \longrightarrow h_{\xi} \xi \Omega^+_{L} = \Omega_R \xi h^+_{\xi},\quad
Q_{L,R} \rightarrow h_{\xi} Q_{L,R}.
\la{nlr}
\ee
Changing to the `dressed' basis implies the following replacements
in the external vector, axial, scalar and pseudoscalar sources
\ba
\bar{V}_\mu & \to & v_\mu = \frac12 \left( \xi^\dagger \partial_\mu \xi -
\partial_\mu \xi  \xi^\dagger +  \xi^\dagger \bar V_\mu \xi +
\xi \bar V_\mu\xi^\dagger - \xi^\dagger \bar A_\mu \xi +
\xi \bar A_\mu \xi^\dagger\right),
\no
\bar{A}_\mu & \to & a_\mu = \frac12
\left( - \xi^\dagger \partial_\mu \xi -
\partial_\mu \xi \xi^\dagger -  \xi^\dagger \bar V_\mu \xi +
\xi \bar V_\mu\xi^\dagger + \xi^\dagger \bar A_\mu \xi +
\xi \bar A_\mu \xi^\dagger\right),\no
\bar{\cal M} & \to & {\cal M} = \xi^\dagger \bar{\cal M} \xi^\dagger .
\la{long}
\ea
Under $SU(2)_L\otimes SU(2)_R$ transformations
\be
v_\mu \rightarrow h_{\xi} v_\mu h^+_{\xi} + h_{\xi} \partial_\mu h^+_{\xi},
\qquad a_\mu \rightarrow h_{\xi} a_\mu h^+_{\xi},\qquad
{\cal M} \rightarrow h_{\xi} {\cal M} h^+_{\xi}. \la{transfor}
\ee
In these variables the relevant part of ECQM action consists of three
 parts
\be
{\cal L}_{ECQM} = {\cal L}_{ch} + {\cal L}_{\cal M} + {\cal L}_{vec},
\la{ECQM}
\ee
where ${\cal L}_{ch}$ accumulates the interaction of chiral fields and
quarks in the chiral limit in the presence of vector and  axial-vector
external fields, 
${\cal L}_{\cal M}$ extends the description 
 for external scalar and pseudoscalar fields and, in particular,
for massive quarks and, finally, ${\cal L}_{vec}$ contains 
operators generating
meson states in vector and axial-vector channels. 

More specifically
\ba
 {\cal L}_{ch} &=&  {\cal L}_0 +
i\bar Q \left( \not\!\! D
 +   M_0 \right) Q
+ i \frac{4 \delta f_0}{\Lambda^2} \bar{Q} a_\mu  a_\mu Q\no
&&+\,\frac{G_{S0}}{4N_{c} \Lambda^2}\,
(\bar{Q}_L Q_R  +
\bar{Q}_R Q_L)^2 - \frac{G_{P1}}{4N_{c} \Lambda^2}\,
( -  \bar{Q}_L \vec\tau Q_R
+  \bar{Q}_R  \vec\tau Q_L)^2,\no
&&+\,\frac{G_{S1}}{4N_{c} \Lambda^2}\,
(\bar{Q}_L\vec\tau Q_R  +
\bar{Q}_R\vec\tau Q_L)^2 - \frac{G_{P0}}{4N_{c} \Lambda^2}\,
( -  \bar{Q}_L  Q_R
+  \bar{Q}_R  Q_L)^2,
\la{ECQM1}
\ea
where
\ba
&& Q \equiv Q_L + Q_R,\no
&&\not\!\! D \equiv \not\!\partial  + \not\! v - \gamma_5 \tilde g_A 
\not\! a , .
\la{va}
\ea
with  the axial coupling $\tilde g_A 
\equiv 1 - \delta g_A$ in the notations of \ci{aet}.  The effective
coefficients appearing in the above expression contain contributions
from modes above $\Lambda_{CSB}$. The term ${\cal L}_0$ contains
operators involving only chiral fields or external vector and
axial-vector sources, such as for instance a term of the form
\be
-\frac{f_0^2}{4}{\rm tr} a_\mu a_\mu .\la{bare}
\ee
We call this and similar pieces `bare' contributions to the
chiral effective lagrangian because they will be renormalized
after integration of the low modes of the quark (and gluon)
fields. Their coupling constants at the scale $\Lambda_{CSB}$ are
expressible
in terms of expectation values of the (high frequency) gluon field operators.
The normalizing constant $\Lambda$ is taken to be $\Lambda_{CSB}$.

As the global chiral symmetry holds under simultaneous transformations of
current quark fields, chiral fields and external sources, the chiral
invariance of the different operators of the lagrangian is certainly
provided in terms of constituent fields. Notice that the couplings
of the four-fermion operators
$G_{S0},G_{S1}, G_{P0}$ and  $G_{P1}$ are in general different.

The massive part ${\cal L}_{\cal M}$ contains the following operators
to the lowest relevant order
\ba
 {\cal L}_{\cal M} &=&
i (\frac12 + \epsilon) \left(\bar Q_R {\cal M}
 Q_L + \bar Q_L  {\cal M}^\dagger  Q_R \right) \no
&& + i (\frac12 - \epsilon) \left( \bar Q_R {\cal M}^\dagger  Q_L
+  \bar Q_L  {\cal M} Q_R\right)\no
&& +  \tr{  c_0\left({\cal M}  + {\cal M}^\dagger\right) 
 + c_5 ({\cal M} +{\cal M}^\dagger)a_\mu a_\mu
 + c_8 \left({\cal M}^2 + \left({\cal M}^\dagger\right)^2\right)}  , \la{massi}
\ea
where $\epsilon$ and $c_0,c_5,c_8$ are real coupling constants.
The couplings $c_0,c_5,c_8$ (which only depend on chiral fields
and external sources) are yet another instance of what we have called
`bare' terms.  The reader will notice that we have changed slightly
our notation with respect to our previous work in \cite{aet}. First
the coefficients $c_i$ are labeled in a way that remind us
of the coefficients of the effective chiral lagrangian
to which they eventually contribute. Second, the matrix
${\cal M}$ introduced here is actually $\xi^\dagger {\cal M} \xi^\dagger$
in the notation of \cite{aet}. The present notation
considerably simplifies our formulae.

Finally, the quark self-interactions in the vector and  axial-vector
channels, ${\cal L}_{vec}$, is implemented by
\ba
{\cal L}_{vec} =&-& \frac{G_{V1}}{4N_c \Lambda^2} \bar Q \vec\tau \gamma_\mu
Q \bar Q \vec\tau \gamma_\mu Q -
\frac{G_{A1}}{4N_c \Lambda^2} \bar Q \vec\tau \gamma_5\gamma_\mu Q
\bar Q \vec\tau \gamma_5\gamma_\mu Q \no
&-&  \frac{G_{V0}}{4N_c \Lambda^2} \bar Q \gamma_\mu Q
\bar Q  \gamma_\mu Q -
\frac{G_{A0}}{4N_c \Lambda^2} \bar Q \gamma_5\gamma_\mu Q
\bar Q  \gamma_5\gamma_\mu Q\no
&+& c_{10} \tr{U \bar L_{\mu\nu} U^\dagger \bar R_{\mu\nu}}. \la{vect}
\ea
where the notations $ \bar L_{\mu\nu}, \bar R_{\mu\nu}$
stand for the strengths of fields $L_\mu, R_\mu$ respectively and, again,
 the couplings
 $G_{V0}, G_{V1}, G_{A0}$ and  $G_{A1}$ are, in general, different.
The term proportional to $c_{10}$ is a `bare' term.

Altogether we have eight $d=6$ four-fermion operators. Of those
we shall only retain those that will be required in our subsequent
analysis and those where a fair comparison with phenomenology
is possible in the $SU(2)$ case. In practice this means
that we retain the operators corresponding to the couplings
$G_{S0}, G_{P1}, G_{V1}, G_{A1}$ which describe the
phenomenology of $I=1$ pseudoscalar, vector and
axial-vector mesons.  The iso-singlet scalar channel is essential
in our analysis.

We complete this section with a remark on the correspondence between
our notations and those ones in a manifestly chiral-symmetric
extended \ci{bij,6,brz} (see also \ci{61})
NJL model with universal scalar, $G_S$, and vector, $G_V$ coupling constants:
\be
G_{S0} = G_{P1} = 4\pi^2 G_S;\qquad G_{V1}= G_{A1} =8\pi^2 G_V.
\ee

A general introduction to chiral effective lagrangian techniques
can be found in \ci{13,131,132}.  Their extension to incorporate
vector and axial-vector states is discussed in
\ci{19a,191,19b,19c}. The derivation of the chiral effective lagrangian
via direct bosonization methods is discussed in
\ci{2,AN,D,aann}.
\section{Bosonization of the ECQM with auxiliary fields}
\hspace*{3ex}Let us add auxiliary fields in the scalar and pseudoscalar
channels, $\widetilde\Sigma, 
\widetilde\Pi \equiv \widetilde\Pi^a \tau^a$, and in the vector and
axial-vector channel,
$\widetilde W^{(\pm)}_\mu =  \widetilde W^{(\pm)a}_\mu \tau^a$, in order to
bilinearize the lagrangian
\gl{ECQM},\gl{ECQM1},\gl{vect} in fermion variables. We replace the
four-fermion interaction by
\ba
 &&\bar{Q}\left[i\widetilde\Sigma  - \gamma_5 \widetilde\Pi
 + \frac12 \gamma_\mu \widetilde W^{(+)}_\mu + 
\frac12 \gamma_\mu \gamma_5 \widetilde W^{(-)}_\mu\right] Q\no
&&+ N_{c} \Lambda^2\left[\frac{\widetilde\Sigma ^2}{G_{S0}} +
\frac{(\widetilde\Pi^a)^2}{G_{P1}} + \frac{
\left(\widetilde W^{(+)a}_\mu\right)^2}{4G_{V1}}
+ \frac{\left(\widetilde W^{(-)a}_\mu\right)^2}{4G_{A1}}\right] \la{HS}
\ea
and include an integration over the
real auxiliary variables $\widetilde\Sigma,
 \widetilde\Pi^a, \widetilde W^{(+)a}_\mu,  W^{(-)a}_\mu$.
After integration of the fermionic degrees of freedom the auxiliary
fields will, generally speaking, propagate. However, some redefinitions
will be required. This is the reason for the tildes in \gl{HS}.

The scalar block of fields in the Dirac operator \gl{exts} reads
\be
\Sigma =M_0 + \widetilde\Sigma +  \frac12 
\left({\cal M} + {\cal M}^\dagger \right)
+\frac{4 \delta f_0}{\Lambda^2} a_\mu a_\mu. \la{sigma}
\ee
Likewise, we group all
pseudoscalar fields in the Dirac action into the pseudoscalar block
\be
\Pi =  \widetilde\Pi + i \epsilon
\left({\cal M}^\dagger - {\cal M} \right). \la{Pi}
\ee
Finally,
the blocks of (antihermitian)
vector and axial vector fields have the
following form
\be
V_\mu = v_\mu - \frac12 i\widetilde W^{(+)}_\mu,\qquad
A_\mu = \tilde g_A a_\mu -  \frac12 i\widetilde W^{(-)}_\mu. \la{VAA}
\ee

The bosonization is completed by integration over the quark fields $\bar Q,
Q$, which induces the quark loop effective action ${\cal W}_{1-loop}$, in
terms of the determinant of \gl{exts}. This determinant must be
regularized with the help of a chirally symmetric regulator \ci{16} and
normalized
at a scale $\mu$. The regulator suppresses frequencies above the cut-off
$\Lambda$, already introduced as an arbitrary scale normalizing the
four-fermion operators, and which is identified with the scale of chiral
symmetry breaking, $\Lambda=\Lambda_{CSB}$. This cut-off is thus
physical and its removal should not be attempted. On the other hand,
$\mu$ is the subtraction point and it is quite arbitrary. The $\mu$
dependence drops from observables, provided that we define the
coefficients of the effective lagrangian ($M_0, G_{S0}, ...$) appropriately
(see \ci{aet}) by introducing running couplings. We may, as we did in
\ci{aet}, choose
the normalization $\mu \simeq \Lambda$. The parameters of the effective
lagrangian are then defined as those at the CSB scale, which simplifies the
expressions noticeably.  In \ci{aet}
a step-function was chosen as regulator. Had we used another regulator,
the result (except for the logarithmically enhanced terms) would have
been different, but so would the `bare' terms in our effective action,
which contain information about higher frequencies. The result would
indeed be scheme dependent. However, since in practice we cannot really
compute these `bare' terms we are limited to using the (universal)
logarithmic terms. This approximation is justified inasmuch as these terms
are numerically dominant.

In particular, the effective potential for a constant $\Sigma
$ field, $\langle \Sigma \rangle \equiv \Sigma_0$,
can then  easily be derived
\be
V_{eff}(\Sigma_0)
= N_c \Biggl\{\frac{ \Lambda^2}{G_{S0}} (\Sigma_0 - M_0 - m_q)^2
+ \frac{1}{8 \pi^2} 
\Sigma_0^4\left( \ln\frac{\Lambda^2}{\Sigma_0^2} + 
\frac12\right)\Biggr\},
\la{pot}
\ee
revealing a
non-trivial minimum given by a solution of the mass-gap equation
\be
\frac{ \Lambda^2}{G_{S0}}\left(\Sigma_0 - M_0 -
m_q\right)
= - \frac{\Sigma^3_0}{4\pi^2} \ln\frac{\Lambda^2}{\Sigma_0^2}. \la{msg2}
\ee
In the weak coupling regime the solution becomes
$\Sigma_0 \simeq M_0 + m_q$
with $ M_0 \gg m_q$ for light $u,d$ quarks. There are no corrections
to this result in the large $N_c$ limit.

It follows from eq.\gl{msg2} that it is natural to work not with
the original parameters $G_{S0}$, $G_{A1}$, etc, but rather
with
\ba
&&\bar G_S = G_{S0}I_0 \frac{\Sigma_0^2}{\Lambda^2},\qquad
\bar G_P = G_{P1}I_0 \frac{\Sigma_0^2}{\Lambda^2}, \no
&&\bar G_V =2 G_{V1}I_0 \frac{\Sigma_0^2}{\Lambda^2},\qquad
\bar G_A = 2 G_{A1}I_0 \frac{\Sigma_0^2}{\Lambda^2},\qquad
I_0 \equiv \frac{1}{4\pi^2}\ln\frac{\Lambda^2}{\Sigma_0^2}. \la{redc}
\ea
The parameters defined with bars are the ones controlling
the departure from the `natural' solution $\Sigma_0 = M_0$.
The weak coupling regime corresponds to
$\bar G \ll 1$. In \ci{aet} we defined similar couplings
but containing $M_0$ rather than $\Sigma_0$. We have found that
our expressions simplify when we use \gl{redc}.

In this notation, and neglecting  higher powers of $m_q$, the solution of
the mass-gap eq.\gl{msg2} reads
\be
\Sigma_0(m_q) \simeq \Sigma_0(0) + m_q 
\frac{1}{1 + 3 \bar G_S};\qquad \Sigma_0(0)\equiv \Sigma_0 
= \frac{M_0}{1 + \bar G_S}. \la{Sig1}
\ee

In practice the
constituent mass is large enough so that a derivative expansion in
inverse powers of $\Sigma_0$ makes sense. We can thus write the
full quark-loop effective action. Retaining the logarithmically enhanced
part we get
\ba
{\cal L}_{1-loop} 
 &\simeq& \frac{N_c}{16\pi^2}  \ln\frac{\Lambda^2}{\Sigma_0^2}\,
 {\rm tr}\, \Bigl(
 (\Sigma^2 + \Pi^2)^2 + (\partial_\mu, \Sigma)^2 + [D^V_\mu, \Pi]^2\no
&&- 4 (A_\mu)^2 \Sigma^2 - \{A_\mu, \Pi\}^2 
- 4i [D^V_\mu, \Pi] \ A_\mu \ \Sigma\ +
2i \partial_\mu \Sigma \{A_\mu, \Pi\} \no
&&- \frac16 \left( (F_{\mu\nu}^L)^2 + (F_{\mu\nu}^R)^2\right)\Bigr) ,
\la{log}
\ea
in terms of \gl{sigma}, \gl{Pi}, and \gl{VAA}. This lagrangian
accumulates the one-loop quark effects and 
together with bare kinetic term for chiral 
fields in \gl{ECQM1} and last lines of \gl{massi}
and \gl{HS} forms the effective meson lagrangian in the presence of external
fields.
\section{ Constant $g_A$, masses and coupling constants of 
vector fields}
\hspace*{3ex}Let us examine the effective lagrangian obtained after
the integration of the quark fields in what concerns the axial-vector
fields. There is a mass term
\be
\Delta {\cal L} =  \frac{N_{c} I_0 \Sigma_0^2}{4}
\tr{\frac{1}{\bar G_{A}}
\left(\widetilde W^{(-)}_\mu\right)^2
+
\left(i 2 \tilde g_A a_\mu +   \widetilde W^{(-)}_\mu\right)^2}. \la{dela}
\ee
This term can be diagonalized \ci{bij,6} by defining
\be
i 2 \tilde g_A a_\mu +   \widetilde W^{(-)}_\mu = 
i 2   g_A a_\mu +   \frac{1}{\lambda_{-}} W^{(-)}_\mu, \la{shif}
\ee
with
\be
 g_A = \frac{\tilde g_A}{1 + \bar G_{A}},\la{g-a}
\ee
which differs from the related expression in the extended NJL model
\ci{bij,6} due to presence of a bare constant $\tilde g_A$.

The constant $\lambda_-$ and its vector counterpart $\lambda_+$ are
determined by requiring the proper normalization of the kinetic term for
physical vector fields $W^{(\pm)}_\mu$.
The appropriate normalization constants
coincide
in \gl{log}
\be
\lambda^2_{+} =\lambda^2_{-} = \frac{N_c I_0}{6}. \la{lVA}
\ee
The masses of vector mesons can be evaluated in the large-log approximation
from \gl{HS} and \gl{log}.
\be
m_V^2
= \frac{6\Sigma_0^2}{\bar G_V}, \la{mV}
\ee
whereas the axial-meson mass is derived from \gl{dela}
\be
m_A^2 = 
6 \Sigma_0^2 \frac{1 + \bar G_A}{\bar G_A} =
 6 \Sigma_0^2 \frac{\tilde g_A}{\tilde g_A -  g_A}. \la{mA}  
\ee
Therefore
\be
g_A = \tilde g_A \frac{m_A^2- 6 \Sigma^2_0}{m_A^2}.
\la{tilde} \ee

Among other characteristics of vector mesons, the coupling constants to
external vector fields are of main importance. They are defined by 
the following term in the lagrangian
\be
\Delta {\cal L} = \frac{i}{4}  \tr{f_V W^{(+)}_{\mu\nu}
\left(\xi \bar L_{\mu\nu}\xi^\dagger + \xi^\dagger \bar R_{\mu\nu} \xi\right)
+ f_A W^{(-)}_{\mu\nu} 
\left(\xi \bar L_{\mu\nu}\xi^\dagger -  \xi^\dagger \bar R_{\mu\nu}
\xi\right)}, \la{tur}\ee
where $W^{(+)}_{\mu\nu}$ and $W^{(-)}_{\mu\nu}$ are the
field strength tensors constructed with $W^{(+)}_{\mu}$
and $W^{(-)}_{\mu}$, respectively. From the previous expression one easily
obtains that
\be
f_V = \lambda_{+};\qquad f_A = g_A \lambda_{-} = g_A f_V.
\ee 
\section{Scalar and pseudoscalar sector: mass spectrum and
decay constants}
\hspace*{3ex}We begin by determining the pion decay constant which can be
found by adding the bare pion kinetic term, \gl{ECQM1}, and the
one obtained from the quark loop, shown in
\gl{dela},  after shifting the fields \gl{shif}
\be
F_0^2 =
f^2_0 + N_c \Sigma^2_0 I_0 g_A \tilde g_A. \la{fpi}
\ee
Recalling that
$a_\mu \simeq 2 \bar A_\mu -i \partial_\mu\pi/F_0 $
the bare kinetic term for pseudoscalar
fields is given by
\ba
\Delta{\cal L} &=& \frac{1}{4}\tr{ (\partial_\mu \widetilde\pi)^2 +
\frac{N_c I_0g_A}{\tilde g_A} 
(\partial_\mu \widetilde\Pi)^2 - \frac{2N_c I_0\Sigma_0 g_A}{F_0}
\partial_\mu \widetilde\Pi \partial_\mu \widetilde\pi} \no
&& +\frac14\tr{m_\pi^2 \widetilde\pi^2 + 
\frac{4 N_c I_0 \Sigma_0^2  \epsilon m_\pi^2}{\bar G_P B_0 F_0}
\widetilde\pi \widetilde\Pi}, \la{kinpi}
\ea
where we have performed a further redefinition
of the $W_\mu^{(-)}$ field so as to cancel the $W_\mu^{(-)}\partial^\mu
\tilde\Pi $ mixing, which brings about another contribution
to the $\tilde\Pi $ kinetic term.

The pion mass is generated by the quark condensate
\be
C_q=i\langle\bar q q\rangle_{eucl} = \left(2 c_0 + \frac{N_c}{4 \pi^2}
\Sigma^3_0 \ln\frac{\Lambda^2}{\Sigma_0^2}\right) \equiv - B_0 F_0^2,
\ee
according to the Gell-Mann-Oakes-Renner formula, $m^2_\pi = 2 m_q B_0$ and
the masses of the $u,d$ quarks are taken equal for simplicity.
The constant $c_0$ (which was named $c_1$ in \ci{aet}) is required
to have a renormalization-group invariant effective potential, as explained
in \ci{aet}.

In the chiral limit, neglecting with the pion mass, one
diagonalizes the kinetic term with the help of the following transformations
\be
\widetilde\pi =  \widetilde\pi' + \frac{\sqrt{1 - \delta^2}}{\delta}
\widetilde\Pi';\qquad
\widetilde\Pi = \frac{1}{\delta} \sqrt{\frac{\widetilde g_A}{g_A N_c I_0}}
\, \widetilde\Pi', \la{diag}
\ee
where we have used the following notation: $\delta \equiv f_0 / F_0$.
As a result the heavy pseudoscalar mass is
\be
m^2_{\Pi} =
\frac{ 2 \Sigma_0^2  \widetilde g_A}{\delta^2 g_A} 
(\frac{1}{\bar G_P} + 1). \la{masPi} 
\ee

In the massless limit one can check that heavy pseudoscalar mesons
completely
decouple from external axial fields and only the pion couples to them
through the vertex $\sim \bar A_\mu \partial_\mu  \widetilde\pi'$.
For massive pions one finds in \gl{kinpi} an additional
mixing between the fields $\widetilde\pi$ and $\widetilde\Pi$.
After a further diagonalization of the mass term (for light pions,
to the first order in
$m^2_\pi$),
\ba
&& \widetilde\pi' \simeq \pi  + d_1 \frac{m^2_\pi}{m^2_\Pi}\Pi;\qquad
\widetilde\Pi'  \simeq - d_1 \frac{m^2_\pi}{m^2_\Pi} \pi + \Pi; \no
&& d_1 =  \frac{\sqrt{1 - \delta^2} }{\delta} \left(
\frac{2 \Sigma_0\epsilon}{\bar G_{P} g_A B_0}
+1\right), \la{d1}
\ea
and the weak decay coupling constant for heavy $\Pi$ meson is found to be
\be
F_{\Pi} = F_0 d_1
\frac{ m^2_\pi}{m^2_\Pi(0)}. \la{fPi}
\ee

The scalar mass is obtained  directly from the lagrangian \gl{HS},
\gl{log} deriving the quadratic form of the fluctuation,
$\Sigma =\Sigma_0 + \tilde\sigma$
\be
m^2_\sigma = 2 \Sigma_0^2 (\frac{1}{\bar G_S} + 3). \la{sigmas}
\ee
The physical scalar field
is given by
$\sigma = \tilde\sigma \sqrt{N_c I_0}$. The reader can easily verify that
all the above formulae reduce to the ones of \ci{aet}
after taking $\tilde{g}_A= g_A$, i.e. $\bar{G}_A=\infty$.

\section{ Chiral symmetry restoration and Weinberg
sum rules}
\hspace*{3ex}It was observed in \ci{aet} that a useful way of constraining the
coefficients of the ECQM was provided by requiring that
at $\mu=\Lambda_{CSB}$ there is an exact matching between
the effective theory and QCD (including both perturbative and
non-perturbative contributions) in those channels which are
sensitive to chiral symmetry breaking. The only example
which was explicitly worked out in our previous work was
the difference between scalar and pseudoscalar
Green functions. In QCD this difference behaves as $1/p^4$, $p^2$
being the squared momentum.

Let us continue this program of exploiting the constraints based
on chiral symmetry restoration at QCD at high energies.
For this purpose we focus on two-point correlators of color-singlet 
quark currents in Euclidean space-time
\ba
\Pi_C (p^2) &=& \int d^4x \,\exp(ipx)\
\langle T\left(\bar q\Gamma q (x) \,\, \bar q \Gamma q
(0)\right)\rangle, \no
&& C \equiv S, P, V, A;\qquad \Gamma = i,\, \gamma_5 \tau^a,\,
\gamma_\mu \tau^a,\,\gamma_\mu \gamma_5 \tau^a.
\ea
In the chiral limit the scalar correlator $\Pi_S$ and the pseudoscalar one 
 $\Pi_P^{aa}$ coincide at all orders
in perturbation theory and also at leading order in the non-perturbative
O.P.E.\ci{11,12,12a}
(see also \ci{aet,ms,19}). In fact
\ba
&&\left(\Pi_P^{aa}(p^2)- \Pi_S(p^2)\right)_{p^2 \rightarrow \infty} \equiv
\frac{\Delta_{SP}}{p^4}  + 
{\cal O} \left(\frac{1}{p^6}\right),\no
&&\Delta_{SP} \simeq  24 \pi\alpha_s C_q^2
\simeq 24 B_0^2 F_0^4,   \la{CSR1}
\ea
where the vacuum dominance hypothesis\ci{11}  has been applied in the
large-$N_c$ limit and the round value
$\alpha_s (1.2 \mbox{\rm GeV}) \simeq 1/3 $ is taken for simplicity\ci{11}.
Therefore the difference \gl{CSR1} represents a genuine order parameter of
CSB in QCD.

The same is true for the difference between the vector, $\Pi_V^{aa}$
and axial-vector, $\Pi_A^{aa}$ correlators \ci{kr,ppr,kpr}
\ba
&&\left(\Pi_V^{aa}(p^2)- \Pi_A^{aa}(p^2)\right)_{p^2 \rightarrow \infty} \equiv
\frac{\Delta_{VA}}{p^6}  + 
{\cal O} \left(\frac{1}{p^8}\right),\no
&&\Delta_{VA} = - 16 \pi\alpha_s C_q^2 \simeq - 16 B_0^2 F_0^4, \la{CSR2}
\ea
where we have defined in the $V,A$ channels
\be
\Pi_{\mu\nu}^{aa} \equiv \left(-\delta_{\mu\nu} p^2 + p_\mu p_\nu\right)
\Pi^{aa}(p^2).
\ee
On the other hand, in the large-$N_c$ limit all correlators are saturated by
narrow resonances \ci{15,151}
\ba
\Pi_P^{aa}(p^2)- \Pi_S(p^2) &=&
\sum_n \,\left[\frac{Z^P_n}{p^2 + m^2_{P,n}} \, -\,
\frac{Z^S_n}{p^2 + m^2_{S,n}}\right],\no
\Pi_V^{aa}(p^2)- \Pi_A^{aa}(p^2) &=&
\sum_n \,\left[\frac{Z^V_n}{p^2 + m^2_{V,n}} \,
-\,\frac{Z^A_n}{p^2 + m^2_{A,n}}\right] - \frac{4 F_0^2}{p^2}.
\ea
As the two above differences decrease rapidly as $p^2$ increases, one can
rightly
expect that only the lowest lying resonances will contribute to (and hence
will be sensitive to)  the constraints from
CSB restoration.

Thus the resonances described by the ECQM; namely a scalar particle, two
pseudoscalar,
a vector and an axial-vector ones, should provide the leading asymptotic 
terms in \gl{CSR1} and \gl{CSR2}. It implies two sets of sum rules
which in the vector channel are
known as the Weinberg Sum Rules \ci{wein}. In particular, in 
the scalar channel one obtains
\ba
&& c_8+\frac{N_c\Sigma_0^2 I_0}{8\bar G_{S}}
-\frac{4\epsilon^2 N_c \Sigma_0^2 I_0}{8\bar G_{P}}=0,\la{srul1}\\
&& Z_\sigma = Z_\pi + Z_\Pi, \qquad  Z_\pi = 4 B_0^2 F_0^2,\la{srul2}\\
&& Z_\sigma  m^2_\sigma =  Z_\Pi  m^2_\Pi + \Delta_{SP}. \la{srul3}
\ea
The first relation determines the bare chiral constant $c_8$
so that to compensate  completely the local 
contributions from four-fermion interaction. Two other ones reduce
the number of independent ECQM coupling constants.

The $\sigma$ and $\Pi$ masses have already been derived in the
previous sections. To impose the chiral symmetry restoration
constraints we need to know the appropriate residues. These can be
obtained from the
part of the ECQM lagrangian which couples scalar and pseudoscalar fields
to their sources. Namely
\ba
\Delta{\cal L}
&=&  \left[- \frac{2N_c \Sigma_0^2 I_0}{\bar G_{S}}\tilde\sigma \bar S
-  \frac{4N_c \Sigma_0^2 I_0 \epsilon}{\bar G_{P}} 
\widetilde\Pi^a \bar P^a
- 2 B_0 F^2_0
\widetilde\pi^a \bar P^a\right]\no
&=& - \alpha \sigma \bar S
- \alpha \beta \pi^a \bar P^a
- \frac{\alpha \left(2 \tilde\epsilon \gamma + \beta \sqrt{1-
\delta^2}\right)}{\delta} \Pi^a \bar P^a ,
\la{Deltal}
\ea
where the last line is written in terms of the
physical fields, i.e. after the diagonalization \gl{diag}.
We have used
the following notation
\be
\alpha \equiv  \frac{2\Sigma_0^2  \sqrt{N_c I_0}}{\bar G_{S}},\qquad
\beta  \equiv \frac{B_0 F_0 \bar G_S}{\sqrt{N_c I_0} \Sigma_0^2},\qquad
\gamma \equiv \frac{\bar G_S}{\bar G_{P}},\qquad
\tilde\epsilon = \sqrt{\frac{\tilde g_A}{g_A}} \, \epsilon.
\la{denot}
\ee 
Comparing the above expressions to the equivalent ones
without vector channels, the only modification consists in the
replacement
$\epsilon \rightarrow \tilde\epsilon$. Therefore all other relations hold 
as well. Namely the sum rules \gl{srul2} and \gl{srul3} lead to
\be
\beta =  
\frac{\sqrt{1 - \frac{m^2_\sigma }{m^2_\Pi}}}{\sqrt{1 - 
\frac{\Delta_{SP}}{Z_\pi m^2_\Pi}}} 
\simeq \sqrt{1 - \frac{m^2_\sigma }{m^2_\Pi}}.\la{cons2}
\ee
The approximation made in the previous expression is justified since
\be
\frac{\Delta_{SP}}{Z_\pi m^2_\Pi} \simeq \frac{6 F_0^2}{m^2_\Pi}
\simeq 0.03. \la{cons3}
\ee
Thus, from phenomenology \ci{10} $|\beta| < 1$.
The relation \gl{srul2} gives
\be
2\tilde\epsilon \gamma  = - \beta\sqrt{1 - \delta^2} \pm  \delta \sqrt{1 -
\beta^2},
\la{beta}
\ee
and, since $|\delta|< 1$, it follows that $|2\tilde\epsilon\gamma|\le 1$.

When taking into account the sum rule \gl{beta}
one can express the parameter $d_1$ in \gl{d1} solely as a
function $\beta$
\be
d_1 = \pm \frac{\sqrt{1 - \beta^2}}{\beta}. \la{d1m}
\ee

The scalar decay constant is
defined through the relation
\be
2 B_0 F_\sigma =\sqrt{Z_\sigma}.
\ee
Therefore
\be
F_\sigma= \frac{\sqrt{N_c I_0}\Sigma_0^2}{B_0 \bar{G}_S}=\frac{F_0}{\beta}.
\la{Fsig}
\ee

It is illustrative to compare the realization of these sum rules in the ECQM
to the one in the usual NJL model, which contains
a (light) massive scalar and a massless pion.
In the corresponding correlators one retains only those poles and therefore
the terms accompanying the constant $Z_\Pi$ should be
dropped. The first sum rule then gives $Z_\pi = Z_\sigma$, i.e.
the characteristic relation of the linear $\sigma$-model. The second sum
rule contains the quantity $Z_\pi m^2_\sigma$. In the NJL model
\be
Z_\pi m^2_\sigma = 16 C_q^2 \frac{\Sigma_0^2}{F_0^2} \simeq  C_q^2
\frac{64\pi^2}{N_c g_A \ln\left(\Lambda^2/\Sigma_0^2\right)}.\la{ineq}
\ee
To derive this result
the well-known
expression for $F_0$ in the NJL model has been used
(equal to our eq. \gl{fpi} with
$f_0=0; \, \tilde g_A = 1$).
Chiral symmetry restoration implies that \gl{ineq} must be equal to
\be
24 \pi\alpha_s C^2_q \simeq  C_q^2
\frac{288\pi^2}{11N_c \ln\left(\Lambda^2/\Sigma_0^2\right)},
\la{ineq2}
\ee
which is obviously not the case since this would require $g_A\simeq
2.5$!
This result is remarkably
independent of colors, flavors and scales.

In the vector channel one derives a similar set of sum rules
\ba
&&c_{10} = 0, \la{vsr1}\\
&&f_V^2 m_V^2 = f_A^2 m_A^2 + F_0^2,\la{vsr2}\\
&&f_V^2 m_V^4 = f_A^2 m_A^4,  \la{vsr3}\\
&&f_V^2 m_V^6 - f_A^2 m_A^6 \simeq  - 4 \pi\alpha_s C_q^2 \simeq - 4 B_0^2
F_0^4, \la{vsr4}
\ea
Eqs.\gl{vsr2}, \gl{vsr3} are the celebrated Weinberg sum rules, while
the last equation \gl{vsr4} was obtained in \ci{kr} (related
considerations can be found also in \ci{kpr},\ci{DG},\ci{Mous}). We
have taken into account that
\be
Z^V_n = 4 f^2_{V,n} m^2_{V,n},\qquad Z^A_n = 4 f^2_{A,n} m^2_{A,n}.
\ee

The first constraint fixes the bare chiral constant $c_{10}$ and implies
the absence of bilinear local operators in the external fields.
The sum rules \gl{vsr2}, \gl{vsr3} determine the
constants $f_V, f_A$ in terms of
vector meson masses and $F_0$. Since in section 3 these same constants have
been determined in terms of $I_0$ and $g_A$ we conclude that
\be
g_A=\frac{m_V^2}{m_A^2}=\frac{\bar G_A}{\bar G_V(1+\bar G_A)},\qquad
f_V^2 = \frac{F_0^2}{m_V^2 (1 - g_A)}
= \frac{N_c I_0}{6},\qquad
f_A^2 = \frac{g_A^2 F_0^2}{m_V^2 (1 - g_A)}. \la{vsum}
\ee

With these constants one can try to saturate the last sum rule \gl{vsr4}
which would imply
\be
F_0^2 m_V^2 m_A^2  \simeq 4 B_0^2 F_0^4, \la{last}
\ee
failing to be satisfied for realistic vector and axial-vector
meson masses and for $B_0 \simeq 1.5$ GeV $\simeq 2 m_V$
\ci{13,21} as $m_A \gg 4 F_0$. 
At this point the lowest-resonance
approximation is not anymore satisfactory which reminds us about the
situation
of the NJL model in the scalar channel.
Certainly  higher-mass, excited vector (and axial-vector)
resonances are needed to correct the asymptotic behavior if \gl{vsr4} is
to be fulfilled.

\section{Chiral constants, fit and discussion}
\hspace*{3ex}We begin this section by obtaining the values of the chiral
constants $L_5$, $L_8$ and $L_{10}$. The derivation of the
first two constants exactly follows the procedure outlined
in \ci{aet}. The results are
\be
L_5 = \frac{N_c \Sigma_0 I_0 g^2_A}{8 B_0(1 + 3 \bar G_S)}.\la{L5}
\ee
\be
L_8 = \frac{1}{64 B_0^2} \left(\frac{Z_\sigma}{m^2_\sigma}
- \frac{Z_\Pi}{m^2_\Pi}\right) =
\frac{F_0^2}{16}\left(\frac{1}{m^2_\sigma} + \frac{1}{m^2_\Pi}\right)
\left(1 - \frac{\Delta_{SP}}{Z_\pi (m^2_\sigma + m^2_\Pi)}\right),
\la{L-8}
\ee
The term proportional to $\Delta_{SP}$ is very small and, in practice,
negligible
\be
\frac{\Delta_{SP}}{Z_\pi (m^2_\sigma + m^2_\Pi)} \simeq
\frac{6 F_0^2}{m^2_\sigma + m^2_\Pi} < 0.03. 
 \la{L-888}
\ee
Thus one expects that
\be
L_8 > \frac{F_0^2}{8 m^2_\Pi}.
\ee

Having considered the vector and axial-vector channels,
we can also estimate the low energy limit of the difference
vector and axial vector correlators
and obtain in this way the value
of the chiral constant $L_{10}$. In the chiral lagrangian, $L_{10}$
parameterizes the operator
\be
- L_{10} \tr{U \bar L_{\mu\nu} U^\dagger \bar R_{\mu\nu}}.
\ee
As $c_{10} = 0$, $L_{10}$ is saturated by vector and axial-vector
exchange. From \gl{tur}
\be
L_{10} = \frac14 \left(f_A^2 - f_V^2\right) = - \frac{F_0^2(1 +g_A)}{4m^2_V}.
\ee

To proceed to an overall fit of the coefficients
of the ECQM and in order to make some physical predictions,
let us first specify the input parameters.
As such we take
$F_0 = 90\ {\rm MeV}$,  $m_\pi^2 = 140\ {\rm MeV}$.Then\ci{13,21}
$\hat m_q (1\ {\rm GeV}) \simeq 6 \ {\rm MeV}$,
$B_0 (1\ {\rm GeV})\simeq 1.5\ {\rm GeV}$,
and, besides, the phenomenological value for the heavy pion mass
$m_\Pi \simeq 1.3$ GeV \ci{10}. We also take the vector meson
masses,
$m_V = m_\rho = 770\ {\rm MeV}$ and  $m_A = m_{a1} \simeq 1.1 \div 1.2$
MeV, as known parameters, though we will not be able to fit
both of them so that to satisfy all sum rules (see discussion in previous
section).

We start by determining the scalar meson mass based on
estimates \ci{13,131,132} for 
the chiral constant $L_8 = (0.9 \pm 0.4)\times 10^{-3}$
\be
m^2_\sigma = m^2_\Pi(
\frac{16 L_8 m^2_\Pi }{ F_0^2} -1)^{-1}\simeq (900 \pm 300) {\rm MeV}. 
\la{msig}
\ee
Employing the mean value of $m_\sigma$, we determine in turn
$\beta \simeq 0.71$.

As discussed, it is not reliable to use \gl{vsr4}, so we will
omit any further reference to it. As for the second
Weinberg sum rule \gl{vsr3}, we will first assume that it holds
(with only one resonance in each channel, that is) to find that
the overall fit is only marginally consistent. Relaxation
of the sum rule \gl{vsr3} will then allow for
a much better fit.

We invoke eq.\gl{vsum} to parameterize
$I_0$ in terms of $g_A$
\be
I_0 = \frac{6 F_0^2}{N_c m_V^2 (1 - g_A)}, \la{I00}
\ee
and eqs.\gl{sigmas} and \gl{denot} to find the dependence of
$\bar G_S$ and $\Sigma_0$ on $g_A$
\be
\bar G_S = \frac{\beta m^2_\sigma}{\sqrt{6}\  B_0 m_V \sqrt{1 - g_A}} -
\frac13, \qquad \Sigma_0^2 = \frac16 m^2_\sigma \left(1 -
\frac{2B_0 m_V \sqrt{1 - g_A}}{\beta m^2_\sigma\sqrt{6}} \right). \la{sig0}
\ee
With the help of these formulae one evaluates $L_5$
\be
L_5 = \frac{F_0^2}{4m_V m_\sigma \beta} \frac{g_A^2}{\sqrt{1 -g_A}}
\sqrt{1 - \frac{2B_0 m_V}{\beta m^2_\sigma\sqrt{6}}\sqrt{1 - g_A}}. \la{l51}
\ee 
Consistency with real values of squared roots implies that
\be
1 > g_A \geq \mbox{\rm min}
\left[1 - \frac{3 m^4_\sigma}{2B_0^2 m_V^2} \left(1 - 
\frac{ m^2_\sigma}{ m^2_\Pi}\right)\right] \simeq 0.48.
\ee
This is the absolute minimum of $g_A$, provided that $\Delta_{SP}$
can be safely neglected. This minimum value of $g_A$ is
attained for $m_\sigma = 1060$ MeV. If we use the central value
$m_\sigma \simeq 900$ MeV one has $g_A \geq 0.60$ for $L_5$ to be
real. However, the latter constant is bounded from the phenomenology
of light pseudoscalar mesons to be $L_5 = (1.4 \pm 0.5)\times 10^{-3}$ for 
sufficiently heavy scalar mesons. Always sticking ourselves to the central
value $m_\sigma \simeq
900$ MeV, the lower value for $L_5$,  $L_5 \simeq 0.9\times 10^{-3}$ is
provided by
$g_A \simeq 0.66$ and the mean value,  $L_5 \simeq 1.4\times 10^{-3}$
is provided by $g_A \simeq 0.71$. 

On the other hand $g_A$ appears in the
relation \gl{vsum} between the masses
of vector and axial vector mesons, $m_A^2 = m_V^2 /g_A$ (that, again,
assumes
that the second Weinberg sum rule is saturated with only one resonance).
Then for
the allowed range of $L_5$ one gets $m_A \leq 0.95$ GeV. Hence, in order to
be as close to the phenomenological value of $m_A = 1.1 \div 1.2$ GeV 
as possible we adopt the lowest possible value for $g_A$ compatible
with the experimentally allowed range for $L_5$ and therefore
$g_A \simeq 0.66$ if $m_\sigma=900$ MeV.
For a heavier scalar meson with $m_\sigma = 1060$ MeV one reaches 
the lowest $L_5$ for  $g_A \simeq 0.62 $ and predicting $m_A \simeq 1$ GeV.

A low value for $g_A$ triggers an unwanted effect, namely,
the diminishing of the dynamical quark mass $\Sigma_0$ as it follows from
\gl{sig0}. To optimize the fit one has to accept larger values for $g_A$
and/or  heavier masses for
scalar meson. In particular, for the mean values of $L_5$ and $m_\sigma$,
i.e. for  $g_A \simeq 0.71$, which leads nevertheless to
unacceptably low values for $m_A$,
one gets  $\Sigma_0 \simeq 150$ MeV.  
The dynamical quark mass is maximal, $\Sigma_0 \simeq 190$ MeV,
for $m_\sigma \sim 1.1$ GeV.

Thus we see that the extension of the ECQM to the
vector and axial-vector channel adds very stringent constraints.
While the overall fit is not bad, it is not brilliant either. In
particular
the tendency to underestimate the axial meson mass is evident and clearly
means that, as already seen when discussing \gl{vsr4},
with only one set of vector and axial vector states
one cannot effectively saturate the Weinberg sum rule \gl{vsr3} either (see
also the hints
in \ci{kr}). Therefore let us accept that only the relation \gl{vsr2} holds
with
a reasonable accuracy and the next one does not. This implies a modification
of eq. \gl{vsum} in the following way
\be
f_V^2 = \frac{F_0^2}{m_V^2 (1 - g_A^2 \xi)}
= \frac{N_c I_0}{6},\qquad
f_A^2 = \frac{g_A^2 F_0^2}{m_V^2 (1 - g_A^2  \xi)},\qquad \xi = 
\frac{m^2_A}{m^2_V} \not= \frac{1}{g_A}  \la{vsum1} 
\ee
The term $\sqrt{1 -g_A}$ in eqs.\gl{sig0},\gl{l51}
should be replaced by  $\sqrt{1 - g_A^2 \xi }$. The parameter $\xi$ takes
the value $\xi \simeq 2.43$ for $m_A \simeq 1.2$ GeV.

Let us perform an optimal fit. For $m_\sigma
\simeq 1$ GeV one finds $\beta \simeq
0.64$ and $L_8 \simeq 0.8\times 10^{-3}$. For $g_A =0.55$ one obtains
$L_5 =  1.2 \times 10^{-3}$ and $\Sigma_0 \simeq 200$ MeV. Then the
vector
and axial vector couplings are $f_V = 0.22$ and $f_A = 0.12 $ to be
compared with the
experimental values \ci{egpr} from  the decay $\rho^0 \rightarrow
e^+e^-, \, f_V \simeq 0.20 \pm 0.01$, and  from
the decay $a_1 \rightarrow \pi \gamma,\,
f_A = 0.10 \pm 0.02$. Then from eqs.\gl{mV},\gl{mA} and \gl{tilde}
one derives that $\bar G_V \simeq 0.25,\, \bar G_A \simeq 0.2, \,
\tilde g_A \simeq 0.66$. In addition $\bar G_S \simeq 0.11$ from
eq.\gl{sigmas}. With these values, $I_0 \simeq 0.1$ and
$\Lambda \simeq 1.3$ GeV. Then the bare pion coupling takes the value $f_0
\simeq 62$ MeV and for the rest of the parameters we find:
$\delta \simeq 0.7$, $\bar G_P \simeq 0.13$,
$\gamma \simeq 0.85$, and either
$\epsilon  \simeq 0.05$ or $\epsilon  \simeq - 0.51$.  The
naive QCD value $\epsilon  \simeq 0.5$ is disfavored.
Finally,
\be
|d_1|= 1.2, \qquad
F_\Pi = 0.8\times 10^{-2} F_\pi, \qquad
F_\sigma = 1.6 F_\pi.
\ee
The constant $F_\Pi$ has been estimated by different methods
 \ci{19b,19c,17,171,18}) to correspond to
$|d_1| = 1 \div 3$. But it is not yet experimentally measured.

\section{ Conclusions}

\noindent
1) In \ci{aet} we established that the QCD effective action that is
appropriate below the scale of chiral symmetry breaking must contain
chiral fields as well as quarks and gluons. The effective action must also
include higher dimensional operators with relatively
weak coupling constants and a relatively large constituent mass. In this
paper we confirm that
the optimal fit to meson physics, including vector and axial-vector channels,
favors weakly coupled four-fermion operators, in contradistinction to
the usual NJL model. The strength of the coupling is always
referred to the constituent mass scale $\Sigma_0$.

\noindent
2) Let us summarize once more the approximations used to
derive the meson characteristics. The  most crucial ones are the large $N_c$
and leading-log approximations. The first one is equivalent to the
neglect of
meson loops. The second one, in fact, is fully compatible with quark
confinement (due to the residual gluon interactions) as
quark-antiquark threshold contributions are suppressed in two-point
functions in the large
log approximation. Furthermore logs are universal and independent of the
method used to separate among low and high frequencies. The accuracy of
this approximation also depends on the influence of heavy
mass resonances which are not included into the particle content of the
ECQM.
One can actually improve of the leading-log approximation with
the help of higher dimensional operators not retained in the ECQM
lagrangian.
But any extension of ECQM should be clearly accompanied with the
opening of new meson channels with physical resonances of higher mass and
spin.

\noindent
3) In this paper we have used a more conventional \ci{13,131,132}
value for
$L_5 = (1.4 \pm 0.5)\times 10^{-3} $ as compared to
$L_5 = 2.2 \times 10^{-3} $ from \ci{22} because the latter one was obtained
in the assumption of a very light scalar meson. However our analysis has made
evident that the low-mass scalar quarkonium bound state should be quite
heavy ($m_\sigma \sim 1$ GeV). Yet it may not be the lightest state as
another
scalar meson  supposedly exists as a $\pi\pi$ scattering state \ci{23} with
a mass  of order $500 - 600$ MeV.

\noindent
4)  The final estimates for effective coupling constants in the scalar and
pseudoscalar channels are shown to be weakly dependent on the presence of
vector mesons and roughly coincide to those ones in \ci{aet}. In particular,
the dominance of two (or the second) mass terms in (11) is again proven.

\noindent
5) The set of CSR constraints predicts unambiguously the value of
axial quark coupling to pions quite a different from unity. It may be close
to 0.5 conjectured in \ci{ppr} but it seems to be unlikely that it
reaches this value due to its being bounded from below because of
$L_5$.

\noindent
6) The predictivity of the ECQM is substantially provided by the CSR
constraints, now enlarged with the Weinberg sum rules. However
only the first of them is well compatible with the particle content of ECQM
while the second and the third ones need at least one more heavy resonance.

\noindent
7) For  completeness we clarify in  more detail the consistency of
the
full set of vector-channel sum rules \gl{vsr2}-\gl{vsr4} in the system with
two vector and one axial-vector resonances. This case was firstly analyzed in
\ci{kr}. A reasonable precision is provided by the fit
\ba
&& m_\rho = 770 {\rm MeV},\qquad m_{a1} = 1170 {\rm MeV},\qquad
m_{\rho'} = 1420 {\rm MeV},\no
&& f_\rho = 0.18,\qquad f_{a1} = 0.11,\qquad f_{\rho'} = 0.05,
 \ea
to be compared to the experimental data \ci{10} and  some phenomenological
estimations \ci{egpr}
\ba
&& m_\rho = 770 {\rm MeV},\qquad m_{a1} = (1230 \pm 40) {\rm MeV},\qquad
m_{\rho'} = (1465 \pm 25) {\rm MeV},\no
&& f_\rho = 0.20 \pm 0.01,\qquad f_{a1} = 0.10 \pm 0.02,
 \ea
as to the value of
$ f_{\rho'}$ it has not been yet measured.
We consider this precision to be excellent for the leading large-$N_c$
approximation (i.e. without any unitarity corrections) and from this we
conclude that the second Weinberg sum rule and the last one \gl{vsr4} indeed
should be applied only to the three-resonance system.

\noindent
8) By adding two more parameters ($G_A$ and $G_V$) we have been
able to predict a large number of physical
parameters in addition to those already determined
in \cite{aet}, such as $f_A$, $f_V$, $m_A$, $m_V$, $L_{10}$ and, in
particular, we get a very clear handle on $g_A$. Perhaps more importantly,
the ECQM appears to be very robust and allows for this
extension without any problems.

\acknowledgments{
This work is supported by EU Network EURODAPHNE,  CICYT grant
AEN98-0431 and CIRIT grant 1998SRG 00026.
A.A. is
supported by the Generalitat de Catalunya (Program PIV 1999) 
and partially by   grants RFFI 98-02-18137 
and GRACENAS 6-19-97. We are grateful to R. Tarrach for useful
discussions and encouragement.}

\end{document}